\documentclass[aps,prl,twocolumn,showpacs,preprintnumbers,amsmath,amssymb,a4paper,superscriptaddress,floatfix,10pt]{revtex4-1}

\usepackage{graphicx}
\usepackage{dcolumn}
\usepackage{bm}
\usepackage{color}

\usepackage{relsize} 

\newcommand*{\pr}{\ensuremath{^\prime}} 

\begin{document}

\title{Current control by resonance decoupling and magnetic focusing in soft-wall billiards}

\author{C. Morfonios}
\affiliation{Zentrum f\"{u}r Optische Quantentechnologien, Universit\"{a}t Hamburg, Luruper Chaussee 149, 22761 Hamburg, Germany}

\author{P. Schmelcher}
\affiliation{Zentrum f\"{u}r Optische Quantentechnologien, Universit\"{a}t Hamburg, Luruper Chaussee 149, 22761 Hamburg, Germany}
\affiliation{The Hamburg Centre for Ultrafast Imaging, Universit\"{a}t Hamburg, Luruper Chaussee 149, 22761 Hamburg, Germany}

\date{\today}

\begin{abstract}
The isolation of energetically persistent scattering pathways from the resonant manifold of an open electron billiard in the deep quantum regime is demonstrated.
This enables efficient conductance switching at varying temperature and Fermi velocity, using a weak magnetic field.
The effect relies on the interplay between magnetic focusing and soft-wall confinement, which rescale the scattering pathways and decouple quasi-bound states from the attached leads, the field-free motion being forwardly collimated.
The mechanism proves robust against billiard shape variations and qualifies as a nanoelectronic current control element.
\end{abstract}


\maketitle

The control of charge flow in low-dimensional quantum systems lies at the heart of nanoelectronic circuit design, posing the challenge to understand and manipulate the mechanisms that enable its realization.
Prominent candidate elements for conductance control are open electron billiards \cite{Zozulenko1997_PhysRevB.56.6931,Nazmitdinov2002,Weingartner2005_PhysRevB.72.115342,Sadreev2006_PhysRevB.73.235342,Racec2010_PhysRevB.82.085313,Rotter2011_PhysRevLett.106.120602}, which can be patterned to almost arbitrary shapes well below the electronic mean free path and coherence length \cite{Fuhrer2001_PhysRevB.63.125309,Heinzel2001_PhysicaE.9.84,Borisov2011_Techn.Phys.Lett.37.136}.
Billiard systems have long served as a convenient platform to study quantum interference phenomena such as Fano resonances \cite{Fano1961,Miroshnichenko2010,Racec2010_PhysRevB.82.085313,Weingartner2005_PhysRevB.72.115342}, but also the quantum-to-classical crossover \cite{Schomerus2004_PhysRevLett.93.154102,Aigner2005_PhysRevLett.94.216801,Rotter2007_PhysRevB.75.125312} and signatures of quantum chaos \cite{Gutzwiller1990_ChaosCQM,Bohigas1984_PhysRevLett.52.1,Jacquod2006_PhysRevB.73.195115}.
Their transport properties are drastically altered by an externally applied magnetic field \cite{Zozoulenko1999_PhysRevLett.83.1838,Rotter2003,Brunner2007_PhysRevLett.98.204101,Buchholz2008,Payette2009_PhysRevLett.102.026808,Morfonios2009,Morfonios2011_PhysRevB.83.205316,Aoki2012_PhysRevLett.108.136804,Kotimaki2013_PhysRevE.88.022913}, and they therefore dominate the intense investigation of coherent magnetotransport in the mesoscopic regime, where quantum interference meets and overlaps with the notion of oriented paths.
Specifically, generalized Aharonov-Bohm (AB) oscillations \cite{Aharonov1959} from phase modulation of interfering states \cite{Sivan1989_PhysRevB.39.1242,Rotter2003,Morfonios2009} combine with the Lorentz deflection \cite{Szafran2005_Europhys.Lett.70.810,Morfonios2011_PhysRevB.83.205316,Poniedzialek2012_JPhysCondMat.24.085801} of electrons up to the formation of edge states \cite{Beenakker1991,Rotter2003,Morfonios2011_PhysRevB.83.205316}.

An intriguing question is how to controllably separate path-mediated magnetotransport dynamics from (resonant) interference effects in a regime where the two strongly overlap, that is, at wavelengths comparable to the system size.
From an experimental viewpoint, the magnetic field provides a unique macroscopic handle on those mesoscopic processes determining the conductance of the system, and the challenge is to find a way to control them under `comfortable' conditions.
In other words:
How can a weak magnetic field switch the current flow through an electron billiard with many levels, at low bias, finite temperature, and over a broad Fermi level variation?
The answer lies in identifying and designing energetically robust transport mechanisms which respond reliably to changes in the field and simultaneously stay isolated from resonance-induced quantum fluctuations.

\begin{figure}[t!]
    \centering
    \includegraphics[width=.88\columnwidth]{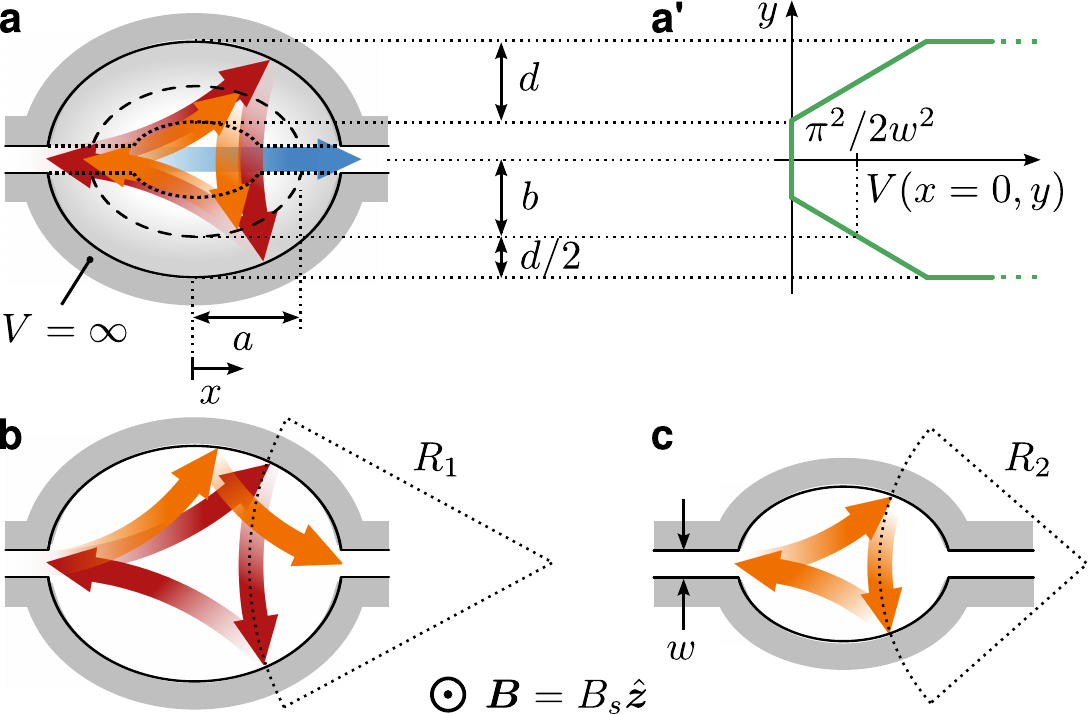}
    \caption{
    (Color online)
    System setup and sketch of pathways. 
    (a) Billiard defined by hard-wall confinement (solid line) and soft-wall potential $V(x,y)$ decreasing along elliptic contours to zero (dotted contour), opening up along $y=0$ to attached leads of width $w$.
    (a\pr) Cross section at $x=0$ for linear wall potential, with central contour at the threshold of the first propagating channel.
    (b) Without the soft wall, a magnetically backscattered pathway (cyclotron radius $R_1$, red arrows) turns into a transmitted pathway ($R_2 < R_1$, orange arrows) for sufficiently lower energy, while (c) backscattering would be retained for a correspondingly smaller billiard.
    For appropriate $V$ in (a), similar backscattered paths can persist in varying energy $E$ for common field $B_s$, with forward propagation favored for $B=0$ (blue arrow).
    }
  \label{fig:fig1}
\end{figure}

In the present work, we realize the above scenario in an open billiard with `soft wall' potential (see Fig.~\ref{fig:fig1}~(a, a\pr)), the experimental setup in mind being a quantum dot with steep boundary potential \cite{Heinzel2001_PhysicaE.9.84,Borisov2011_Techn.Phys.Lett.37.136} supplied with additional peripheral gates \cite{Fuhrer2001_PhysRevB.63.125309}.
The combination of elongated lateral shape and soft wall is shown to control magnetotransport by isolating required scattering pathways from resonant levels.
For $B=0$, the incoming electrons are directed forwardly causing high transmission, whereas at a switching field $B=B_s$ they are focused \cite{Beenakker1991} into a completely backscattered pathway which becomes geometrically `rescaled' in energy (see Fig.~\ref{fig:fig1}~(a)).
In both cases, the crucial role of the soft wall is to create energetically persistent scattering pathways while decoupling quasi-bound states from the openings.
As a result, the setup enables efficient finite-temperature current switching via a weak magnetic field, for varying Fermi energy.

\begin{figure}[t!]
    \centering
    \includegraphics[width=0.98\columnwidth]{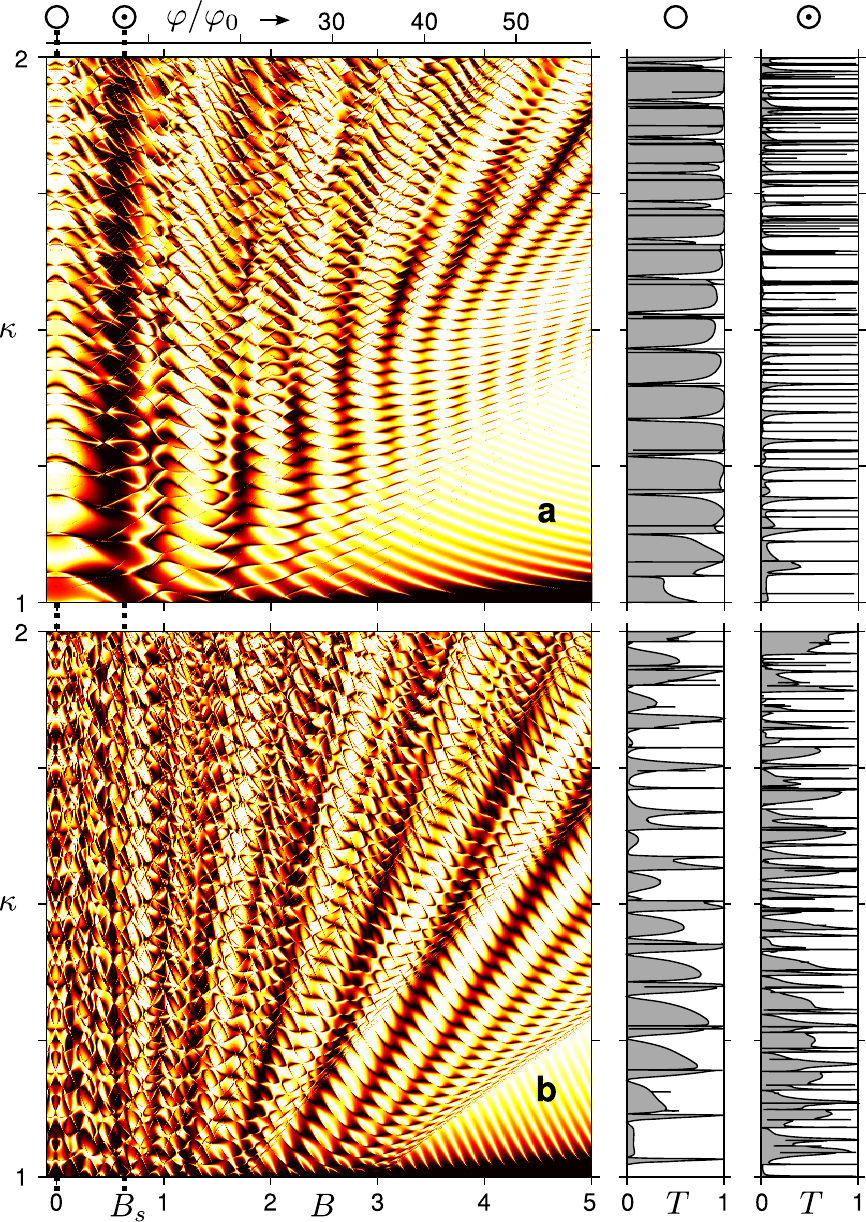}
    \caption{
    (Color online)
    Transmission $T$ (black$=0$ to white$=1$) for varying magnetic field $B$ (or flux $\varphi$) and scaled incoming momentum $\kappa = \sqrt{2E}w/\pi$ within the first open channel, for (a) the soft-wall potential of Fig.~\ref{fig:fig1}~(a\pr) with $(a,b,d,w)=(84,128,96,32)~a_0$, and (b) the same billiard without soft wall.
    $B$ is in units of $10^{-3}~B_0$. For $a_0 = 2~{\rm nm}$, $B_0 = \hbar /ea_0^2 = 164.55~\rm{T}$.
    Right panels: cuts through $T(B,\kappa)$-maps at $B=0$ ($\mathlarger{\mathsmaller{\mathsmaller{\bigcirc}}}$) and $B_s=0.63\times 10^{-3}~B_0$ ($\odot$) or $\varphi_s = 7.32~\varphi_0$ (flux quantum $\varphi_0 = h/e$).
    }
  \label{fig:fig2}
\end{figure}

With decohering electrodes implemented by attached semi-infinite leads, the effective (energy dependent and non-Hermitian \cite{Datta1995}) Hamiltonian of the open system is represented on a tight-binding lattice, and the transmission function $T(E)$ is computed via an extended recursive Green function scheme \cite{FerryGoodnick1997,Drouvelis2006,Morfonios2011_PhysRevB.83.205316}.
This allows for efficient and accurate transport calculations in a highly resolved parameter space for the considered low-energy regime.
The conductance $G$ at Fermi energy $E_F$ and temperature $\varTheta$ is then obtained from $T(E)$ within the Landauer-B\"uttiker framework \cite{Datta1995}.
Upon an excitation in the leads, the Green function further provides the local density of states (LDOS) $\rho$ as well as the scattering wave function \cite{Datta1995} which in turn yields the probability current density $\bm j$ \cite{Baranger1991_PhysRevB.44.10637,Zozoulenko1996_PhysRevB.53.7975}.
The choice $\hbar = e = m = a_0 \equiv 1$ fixes the units of energy $E_0 = \hbar^2 / m a_0^2$ and magnetic field strength $B_0 = \hbar / ea_0^2$ for given effective mass $m$ and lattice constant $a_0$.

\begin{figure*}[t!]
  \centering
  \includegraphics[width=.98\textwidth]{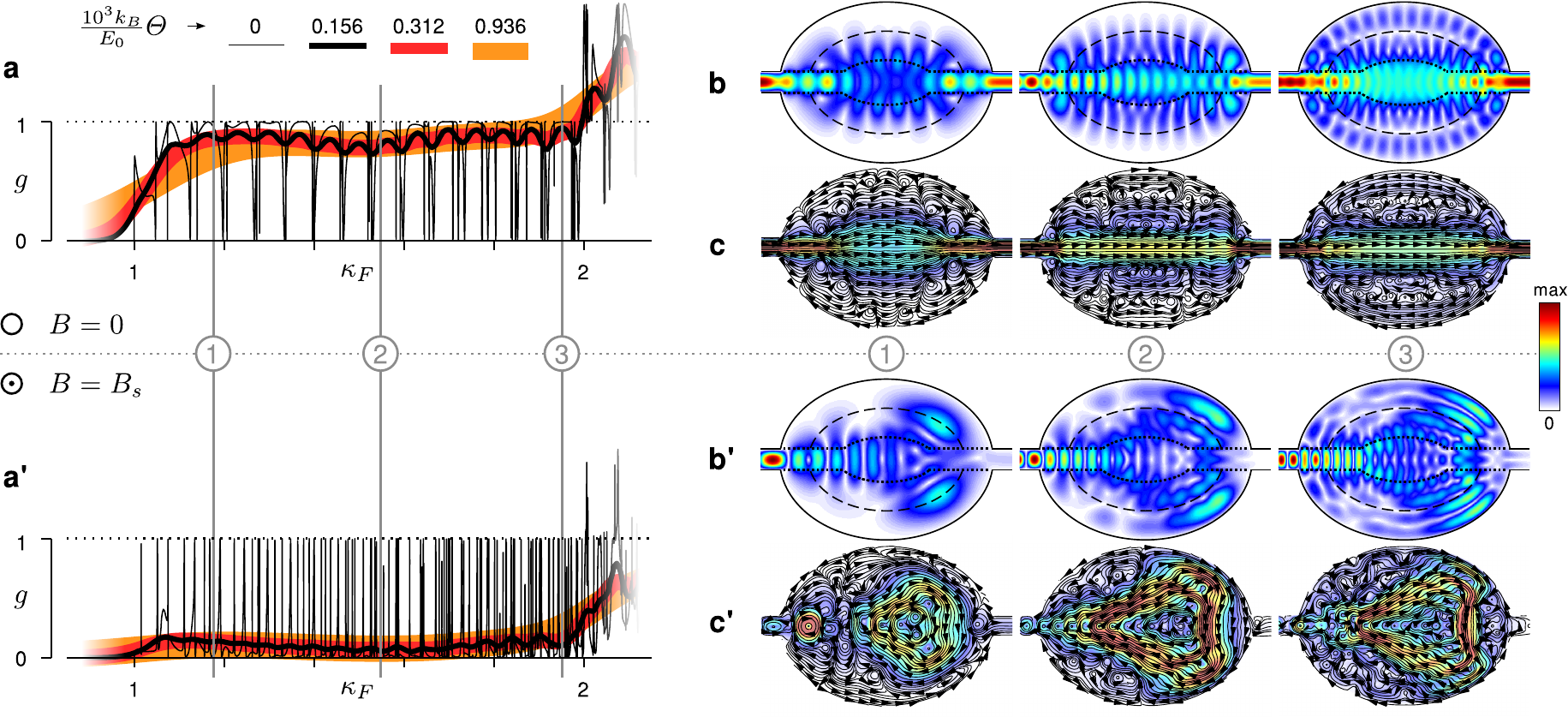}
  \caption{
  (Color online)
  (a) Dimensionless conductance $g = G/G_0$ (with quantum $G_0 = 2e^2/h$) around the first open channel for $B = 0$, for the same billiard as in Fig.~\ref{fig:fig2}~(a), at different temperatures $\varTheta$. 
  For $a_0 = 2~{\rm nm}$ and $m = 0.069~ m_e$ (GaAs/AlGaAs interface): $E_0 = 276~\rm{meV}$ and $\varTheta = 0, 0.5, 1.0, 3.0~{\rm K}$.
  Scaled (b) LDOS $\sqrt{\rho}$ and (c) current density $\sqrt{|\bm{j}|}$ shown at momenta indicated by vertical lines in (a), for electrons incident in the left lead.
  (a\pr, b\pr, c\pr) Same as above, but for $B = B_s$.
  }
  \label{fig:fig3}
\end{figure*}

The transmission through the billiard with and without soft-wall is shown in Fig.~\ref{fig:fig2} for varying $B$ and incoming momentum $\kappa$.
Qualitatively common features in the two $T(B,\kappa)$-maps are: 
isolated edge state peaks \cite{Beenakker1991} at high $B$ and low $\kappa$ (lower right corner) as well as stripe-like interference patterns from multiple edge states \cite{Rotter2003} for higher $\kappa$ (lower diagonal half), which become less pronounced and eventually destroyed by generalized AB interference of spatially extended states \cite{Sivan1989_PhysRevB.39.1242,Morfonios2009} at lower $B$ (upper diagonal half).
The slope of the characteristic reflection and transmission stripes portrays the formation of skipping \cite{Beenakker1991} orbits within the billiard.
Without the soft wall (Fig.~\ref{fig:fig2}~(b)), the approximate commensurability between the skipping intervals and the (half) length of the boundary is preserved along stripes of positive slope in the $(B,\kappa)$-plane on which high reflection (transmission) occurs.
The soft wall causes the stripes to bend around the middle of the channel (Fig.~\ref{fig:fig2}~(a)), which shows that the finite potential effectively reduces the size of the billiard area at low $\kappa$:
A stronger focusing field is needed to maintain the high or low $T$ for decreasing $\kappa$, and consequently the map features broaden along the $B$-axis.

In the present context, the extraordinary effect of the soft wall is manifest in the low-$B$ regime, where the very complex dynamics generally induces highly irregular interference contributions:
At a relatively weak field $B = B_s$, backscattering persists over the whole channel and  completely dominates the background transmission spectrum, forming a broad and vertical reflection stripe in the $T(B,\kappa)$-map, upon which only very narrow Fano resonances are superimposed (see Fig.~\ref{fig:fig2}~(a)).
Indeed, background transmission does not set in again until the second channel threshold.
This remarkable feature, which is absent without the soft wall (see Fig.~\ref{fig:fig2}~(b)), is reversed when the field is turned off:
For $B = 0$, a highly transmitting background is only slightly perturbed by narrow resonant dips.
The cuts through the $T(B,\kappa)$-maps in Fig.~\ref{fig:fig2} highlight the above behavior.
At finite temperature, the dips (peaks) at $B = 0$ ($B = B_s$) are effectively washed away by the thermal contribution of the highly transmitting (reflecting) non-resonant states around the Fermi level.
This is seen in Fig.~\ref{fig:fig3}~(a) or (a\pr), where the conductance is kept close to unity or zero, respectively, over a broad range in $E_F$ even at considerable thermal width $k_B\varTheta$.

To understand the influence of the proposed type of soft wall potential, and the induced mechanism underlying conductance control, let us analyze the electronic scattering states responsible for high (low) background transmission in the absence (presence) of the field.
Fig.~\ref{fig:fig3}~(b, b\pr) displays the LDOS $\rho(x,y;\kappa)$ for electrons incident in the left lead of the billiard at sample non-resonant energies.

For $B = 0$ (Fig.~\ref{fig:fig3}~(b)), we see that the effect of the finite potential is to direct the motion along the axis connecting the leads, thus enhancing transmission.
This is achieved in a twofold way:
(i) The special shape of the potential around the lead openings, forming a stub of free motion into the billiard as a prolongation of each lead, suppresses the transversal component of the electronic local momentum, thereby collimating \cite{Beenakker1991} the motion in forward direction (in other words, the soft wall reduces the diffractive effect of the hard-wall lead openings).
(ii) Owing to its elliptic contour, the soft wall depletes the scattering state along the billiard boundary and further confines it into an elongated profile leaking into both leads.
For the same reason, states corresponding to distinct Fano resonances become well decoupled from the leads, and thus isolated from a significant (subtractive) contribution to the overall transport.

For $B = B_s$ (Fig.~\ref{fig:fig3}~(b\pr)), the scattering state profiles reveal the key role of the soft wall in energetically sustaining the backscattered pathways.
Again, the mechanism is twofold:
(i) States strongly coupled to the incoming lead are now magnetically focused onto the billiard boundary, so that the electron follows a pathway which is backscattered after `bouncing' twice off the boundary \cite{Brunner2007_PhysRevLett.98.204101,Ferry2011_Semicond.Sci.Technol.26.043001}.
The soft wall here crucially comes to the aid of conductance suppression by rescaling the dynamics and thus keeping the non-resonant backscattered pathway energetically invariant:
With increasing (decreasing) kinetic energy, the electron undergoes weaker (stronger) Lorentz deflection at constant $B=B_s$, but at the same time penetrates more (less) into the soft wall potential towards the boundary (compare outer lobes of $\rho$ in Fig.~\ref{fig:fig3}~(b\pr; 1,2,3)).
The soft wall thus effectively increases the billiard size with energy, and as a result, the magnetically focused, backscattered pathway persists over the whole channel.
(ii) As in the field-free case, any long-lived resonant states are further confined away from both leads by the soft wall, rendering the corresponding Fano peaks extremely narrow.

The actual electronic motion in the billiard is depicted in Fig.~\ref{fig:fig3}~(c, c\pr)) through its probability current density $\bm{j}(x,y;\kappa)$.
With or without magnetic field, the wave nature of transport leads to multiple complex vortex structures covering the billiard, which change dramatically in energy.
Nevertheless, we see that the parts of the flow with higher density indeed favor motion along the above described pathways needed for conductance switching in varying $E_F$, that is, a forward collimated current for $B = 0$ and a circulating backscattered current for $B = B_s$.

It should be pointed out that, although the soft wall succeeds in geometrically rescaling the low-field (two-bounce) backscattered pathway, the motion is in general drastically modified from that in a corresponding purely hard-wall billiard with spectral boundary reflection \cite{Aoki2012_PhysRevLett.108.136804}.
In the present case, the further into the soft wall the electron reaches, the more it is magnetically deflected due to its reduced (local) momentum, and the motion is further affected by continuous electrostatic refraction \cite{Repp2004_PhysRevLett.92.036803}.
These effects are enhanced at stronger fields which localize the scattering states closer to the boundary over longer parts (unlike the two-bounce paths, which predominantly enter the wall radially).
Therefore, such higher order (four-bounce, six-bounce, etc.) backscattered pathways \cite{Brunner2007_PhysRevLett.98.204101,Brunner2012_J.Phys.Cond.Mat.24.343202} cannot persist over large energy intervals for the same potential.
Indeed, in Fig.~\ref{fig:fig2}~(a) vertical reflection stripes tend to form also at higher field strengths ($B/B_0 \approx 1.8, 2.2$, etc.), but are eventually tilted or destroyed as energy varies.
Switching efficiency is thus restricted to smaller $\kappa_F$-range and lower $\varTheta$ at these fields.

\begin{figure}[t!]
    \centering
    \includegraphics[width=0.98\columnwidth]{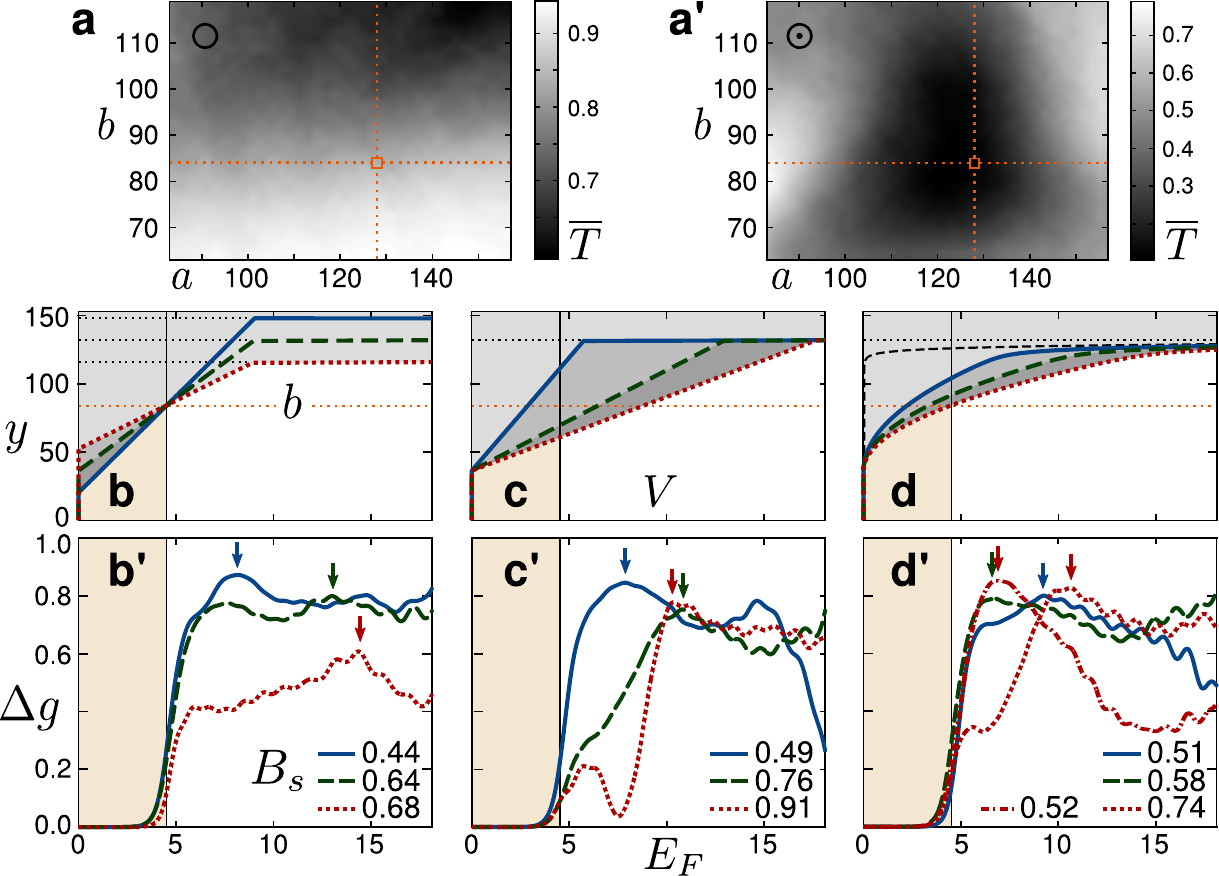}
    \caption{
    (Color online)
    (a, a\pr): 
    Channel-averaged transmission for varying mid-wall semi-axes $a$ and $b$ of the billiard in Fig.~\ref{fig:fig1} with $d=96~a_0$, at (a) $B=0$ and (a\pr) $B=B_s$.
    Dotted lines indicate the geometry $(a,b)=(128,84)~a_0$ chosen in Figs.~\ref{fig:fig2},~\ref{fig:fig3} and below.
    (b, c, d):
    Cross section $V(x=0,y)$ for different (b, c) linear and (d) parabolic (with Wood-Saxon-type \cite{Betancur1998_JPhysD.31.3391} boundary, thin dashed line) soft wall profiles.
    (b\pr, c\pr, d\pr): Corresponding switching contrast $\Delta g$ at $k_B\varTheta/E_0 = 0.312\times10^{-3}$ ($\varTheta = 1.0~{\rm K}$ for $a_0 = 2~{\rm nm}$, $m = 0.069~ m_e$) for optimal fields $B_s$ within the first open channel (threshold at vertical lines).
    Dash-dotted line in (d\pr) corresponds to dotted profile in (d) for alternative $B_s$.
    Arrows indicate each curve's maximum.
    Lengths, energies and fields are in units of $a_0$, $10^{-3}E_0$ and $10^{-3}B_0$, respectively.
    }
  \label{fig:fig4}
\end{figure}

Having demonstrated and explained the proposed switching mechanism, we finally analyze the impact of setup variations.
In Fig.~\ref{fig:fig4}~(a, a\pr), the channel-averaged transmission $\overline{T} = \int_1^2 T(\kappa)\,d\kappa$, a simple estimate of the overall transmittivity, is shown for varying lateral shape of the billiard.
The substantial overlap between high-$\overline{T}(B=0)$ and low-$\overline{T}(B=B_s)$ areas indicates the robustness of the switching effect against alteration of the dot shape.
For a chosen shape, Fig.~\ref{fig:fig4}~(b\pr, c\pr, d\pr) shows the switching contrast $\Delta g = g(B=0)-g(B=B_s)$ at a reference broadening $k_B\varTheta$ for different soft wall profiles, including ones (d) that simulate a concrete experimental setup \cite{Fuhrer2001_PhysRevB.63.125309,Heinzel2001_PhysicaE.9.84,Borisov2011_Techn.Phys.Lett.37.136}.
Note that the high switching efficiency relies on the enhanced $g(B=0)$ and suppressed $g(B=B_s)$ of a single and relatively large billiard (of area $\gg w^2$) containing many resonant levels ($>130$ within the first channel at $B=0$) isolated from the leads, and is achieved for a broad variety of soft wall profiles at substantial thermal width \cite{note}.
The optimal switching field $B_s$ generally increases with the steepness of the wall potential, in accordance with the stronger confinement of low-energy states.
Further, optimal switching (maximal $\Delta g$, see arrows in Fig.~\ref{fig:fig4}~(b\pr, c\pr, d\pr)) can be adjusted to different $E_F$ by changing the soft wall parameters.
For certain setups (dotted line in (d), corresponding to dotted and dash-dotted lines in (d\pr)), energy-persistent backscattering (large $\Delta g$) occurs for distinct $B_s$-values along separate parts of the channel, meaning that optimal $E_F$ for switching can be magnetically tuned in this case.

The experimental realization of the proposed switching device is feasible, e.g., in Ga[Al]As heterostructures by a combination of local oxidation techniques with optical or electron-beam lithography \cite{Fuhrer2001_PhysRevB.63.125309,Heinzel2001_PhysicaE.9.84,Borisov2011_Techn.Phys.Lett.37.136}.
This provides a high precision in lateral dot shape with steep soft-wall potential corresponding to a depletion length $\sim 15~{\rm nm}$ \cite{Heinzel2001_PhysicaE.9.84}.
The quantum dot can be tuned by additional top or planar gates \cite{Fuhrer2001_PhysRevB.63.125309,Heinzel2001_PhysicaE.9.84}, and large electron mean free paths are achievable at low temperature (e.g., $3$--$5~\mu{\rm m}$ at $4.2$~K \cite{Borisov2011_Techn.Phys.Lett.37.136}), which is important in order to maintain as high degree of ballisticity as possible \cite{See2012_PhysRevLett.108.196807}.
Since the proposed switching device consists of a single dot, its fabrication is also facilitated below the electronic coherence length above $\varTheta\sim 1$~K \cite{Beenakker1991,Fuhrer2001_PhysRevB.63.125309}.
Even in the presence of (weak) dephasing, though, the desired switching effect should in fact be enhanced, since it relies on the suppression of resonant interference:
In similarity to the thermal averaging taken into account, dephasing would attenuate the Fano extrema \cite{Baernthaler2010_PhysRevLett.105.056801} and thus contribute to the overall high versus low conductance profile needed for robust switching.

In conclusion, we have demonstrated how to isolate the magnetically controllable scattering continuum from the manifold of resonant levels of a many-level electron billiard, persistently in energy.
The underlying mechanism relies on the combined action of an elongated (elliptic) billiard boundary and a designed soft-wall potential, which together decouple quasi-bound states from the attached leads while simultaneously directing forward field-free transport or geometrically rescaling magnetically deflected, backscattered paths.
The proposed setup constitutes an efficient and robust conductance switching device operating at finite temperature, weak magnetic field and over broad Fermi level variation, and is realizable with current experimental techniques.

The authors are thankful to P. Giannakeas for valuable discussions.

\end{document}